\documentclass[12pt]{iopart}

\usepackage{iopams}

\newcommand{\bk}{{\bm{k}}}

\newcommand{\acm}[2]{\{#1,#2\}}

\usepackage{ulem}
\usepackage[]{graphicx}
\usepackage{xcolor}
\usepackage[]{amssymb}
\usepackage[]{amsfonts}
\usepackage[]{latexsym}
\usepackage[]{float}
\usepackage[all]{xy}
\usepackage{bm}

\newcommand{\pics}[0]{\,}
\newcommand{\PFmat}[1]{\pics\nm{#1}{24}{24}}
\newcommand{\PHmat}[1]{\pics\nm{#1}{15}{15}}
\newcommand{\Pket}[1]{\pics\nm{#1}{15}{24}}
\newcommand{\PTket}[1]{\pics\nm{#1}{24}{15}}

\newcommand{\PFvC}[1]{\pics\nm{#1}{5}{24}}

\newcommand{\nm}[3]{
\begin{xy}
  0*=<#2mm,#3mm>[F]{
    { #1}}="box";
"box"
,
"box"
\end{xy}}

\newcommand{\eas}[0]{\begin{eqnarray*}}
\newcommand{\eae}[0]{\end{eqnarray*}}
\newcommand{\les}[0]{\begin{equation}}
\newcommand{\lee}[0]{\end{equation}}
\newcommand{\leas}[0]{\begin{eqnarray}}
\newcommand{\leae}[0]{\end{eqnarray}}
\newcommand{\mchss}[4]
{
\left\{
\begin{array}{cc}
#1 & #2   \\
#3 & #4
\end{array}
\right.
}

\newcommand{\mmat}[4]
{
\left[
\begin{array}{cc}
#1 & #2 \\
#3 & #4 
\end{array}
\right]
}

\newcommand{\mvec}[2]
{
\left[
\begin{array}{c}
#1  \\
#2  
\end{array}
\right]
}

\newcommand{\mvecthree}[3]
{
  \left[
    \begin{array}{c}
#1  \\
#2  \\
#3  
    \end{array}
    \right]
}
\newcommand{\mvecfour}[4]
{
\left[
\begin{array}{c}
#1  \\
#2  \\
#3  \\
#4  
\end{array}
\right]
}

\begin{document}

\title[
  Flat bands in Weaire-Thorpe model and silicene]
{  
  Flat bands in Weaire-Thorpe model and silicene 
}

\author{Y. Hatsugai}
\address{
Division of Physics and Tsukuba Research Center for Interdisciplinary Materials Science,
University of Tsukuba,
Ibaraki 305-8571, Japan
}
\address{
  Center for Innovative Integrated Electronic Systems,
   Tohoku University, 
  Sendai, Miyagi 980-0845, Japan
}
\ead{hatsugai@rhodia.ph.tsukuba.ac.jp}

\author{K. Shiraishi}

\address{
 Department of Computational Science and Engineering,
Nagoya University,
Nagoya 464-8603, Japan
}
\address{
  Center for Innovative Integrated Electronic Systems,
   Tohoku University, 
  Sendai, Miyagi 980-0845, Japan
}

\author{H. Aoki}

\address{
Department of Physics, University of Tokyo, 
Tokyo 113-0033 Japan }

\begin{abstract}
In order to analytically capture and identify peculiarities in the 
electronic structure of silicene, 
Weaire-Thorpe(WT) model, a standard model for treating 
three-dimensional (3D) silicon, is applied to silicene with 
the buckled 2D structure.  In the original WT model for 
four hybridized $sp^3$ orbitals on each atom along with inter-atom hopping, 
the band structure can be systematically examined in 3D, where 
flat (dispersionless) bands exist as well.   
For examining silicene, here we re-formulate the WT model 
in terms of the  overlapping molecular-orbital (MO) method  
which 
enables us to describe flat bands away from the electron-hole
symmetric point. 
The overlapping MO formalism indeed 
enables us to  reveal an important difference: while 
in 3D the dipersive bands with cones are sandwiched 
by doubly-degenerate flat bands, in 2D 
the dipersive bands with cones are sandwiched 
by triply-degenerate and non-degenerate (nearly) flat bands, 
which is consistent with the original band calculation 
by Takeda and Shiraishi.   
Thus emerges a picture for why the whole band structure 
of silicene comprises a pair of dispersive bands with Dirac cones 
with each of the band touching a nearly flat (narrow) band at $\Gamma$. 
We can also recognize that, for band engineering, the bonds perpendicular to the atomic plane are crucial, and that 
a ferromagnetism or structural instabilities  are  expected if we can 
shift the chemical potential close to the flat bands.
  \end{abstract}

\maketitle

\section{Introduction}
After the physics of graphene was kicked off,
originally by a 
  theoretical prediction for a 
massless Dirac fermion by Wallace\cite{Wa} back in the 1950's 
and by a recent experimental realization\cite{GN}, 
interests are extended to wider class of systems.  
Unusual behaviours of the
massless Dirac fermions are then experimentally confirmed 
and theoretically analysed\cite{Ao}. 
History seems to repeat itself, for silicene: the material was
theoretically predicted 
in the early 1990's by one of the present authors
and Takeda\cite{TS},
and the material began to be synthesised recently\cite{Lalmi,Vogt,Lin},
after such a long latent period.
Successful synthesis of silicene
on Ag 
has triggered 
intense attentions in both physics and material science.  
In graphene, the carriers, being in a atomically flat 
system, are $\pi$-electrons arising from 
the $sp^2$ hybridization 
of carbon orbitals.  In silicene, by contrast, 
the honeycomb lattice is buckled, so that 
a $sp^3$ character of Si is involved, which 
is crucial as pointed out already in \cite{TS}.  
Thus silicene is not just a Si analogue of graphene, 
but distinct multi-orbital characters should 
appear in its electronic structure.

Hence silicene has a larger degree of freedom than 
graphene as a target material 
for applications and also theoretical considerations.  
In the present paper, we focus on this multi-orbital 
feature of silicene, where we opt for a simplified
model, namely we propose to introduce 
as an extension of the  Weaire-Thorpe(WT) model, which 
was originally conceived
for a three-dimensional (3D) silicon with $sp^3$ orbits in a tight-binding
model on a diamond lattice\cite{WT}.  
In the original WT model for 
four $sp^3$ orbitals on each atom along with inter-atom hopping in 3D, 
the band structure is systematically examined, and 
two singular dispersions arise: 
One is a massless Dirac cone, and the other is dispersionless 
(flat) band\cite{DF86}. 
{
Massless Dirac cones in three dimensions are 
topologically protected.  
In two dimensions, they can arise, with supplemental symmetry protections, 
as in graphene and silicene.\cite{HAspringer}.  
On the other hand, the flat bands are not protected by symmetry in realistic materials. Still, however,
 the flat bands can have some topological/geometrical
 origins reflecting the multi-orbital character of a given material
 as we describe in the present paper.}  
For silicene, here we re-formulate the WT model 
in terms of the overlapping molecular-orbital (MO) method\cite{ZQ} 
  discussed by Hatsugai and Maruyama, 
which contains WT and enables us to describe flat bands away from the electron-hole symmetric point.  
The overlapping MO formulation indeed 
enables us to pin point, algebraically, an important difference: while 
in 3D the dipersive bands with cones are sandwiched 
by doubly-degenerate flat bands, in 2D 
the dipersive bands with cones are sandwiched 
by triply-degenerate and non-degenerate (nearly) flat bands, 
which is consistent with the original band calculation 
by Takeda and Shiraishi\cite{TS}.   
Thus emerges a picture for why the whole band structure 
of silicene comprises dispersive bands with Dirac cones 
along with nearly flat (narrow) bands.

In the present paper, we first start with an 
{\it overlapping molecular orbital theory}\cite{ZQ}.  
Applying this to the WT model enables us to generically treat the flat bands away from the electron-hole symmetric 
point in multi-orbital models
while the usual flat-band theories\cite{ST,Lieb,ShimaAoki,HAnjp} focus on those 
at the electron-hole symmetric point. 
We then gives a picture for silicene in this formalism, and 
compare the result with the band structure due to 
Takeda and Shiraishi.   
We finally discuss that 
the nearly flat bands imply large density of states (DOS), 
which will give an interesting possibility for inducing 
instabilities into symmetry-broken 
states such as ferromagnetism or structural changes.  
For the band engineering the bonds perpendicular to the atomic plane are suggested to be crucial.

\section{Overlapping molecular orbitals and flat bands}
\label{sec:overlap}

Let us start with describing a class of lattice model Hamiltonians in terms of molecular orbitals proposed by 
Hatsugai and Maruyama\cite{ZQ}, which is here 
explained to make the present paper self-contained. 
Consider fermions on an $N$-site lattice 
with the creation operator, 
$c_i^\dagger$, for the sites $i=1,\cdots,N$ with $\acm{c_i^\dagger}{c_j}=\delta _{ij}$.  
After shifting the origin of energy by $\mu $, let us
consider the case in which the Hamiltonian is 
given as  a sum of
  {\it overlapping molecular orbitals} $m (=1, ..., M)$ as
\begin{eqnarray*}
  H -\mu {{\cal N} } =  \sum_{m=1}^M {\cal E}_m C_m ^\dagger C_m , 
  \end{eqnarray*}
where ${\cal N}  =  \sum_{i=1}^N c_i ^\dagger c_i$
is the number operator of the fermions.
The coefficient ${\cal E}_m\in\mathbb{R}$ is 
the energy level of the molecular orbital $m$, while 
$C_m ^\dagger $ creates the molecular orbital  as
\begin{eqnarray*} 
  C_m ^\dagger  &=& \sum_{i=1}^N c_i ^\dagger \psi_{i,m} = {c} ^\dagger {\psi}_m,\ 
  {c} ^\dagger =(c_1 ^\dagger ,\cdots,c_N ^\dagger ), \ {\psi}_m =\mvecthree{\psi_{1,m}} {\vdots} {\psi_{N,m}} ,
\end{eqnarray*}
      where $\psi_m$ is the wavefunction of the molecular orbital $m$.  
 We do {\it  not}  require translational symmetry in the 
  system or in the molecular orbitals 
  as is the case with the WT model.

Then we have a simple theorem that 
$H-\mu {\cal N}$  has $(N-M)$-fold degenerate zero-energy states 
when $N-M>0$, 
where $M$ is the total number of the molecular orbitals
in the whole real space.  
  The number of the zero energy states
  $N-M$ can be macroscopic.
 Since the theorem is general, we can apply it to 
random systems but to periodic systems as well.
Then we can characterise the wavefunctions as Bloch 
states, and $N$ may be regarded as the number of 
energy bands when we express the Hamiltonian 
in the Bloch basis.
The total number of the molecular orbitals, $M$,
in this Bloch basis corresponds to the total 
number of the terms in the Hamiltonian in the 
momentum representation, which varies 
from a model to another, as we shall see.

Note here that if there are zero-energy states
in the present Hamiltonian, they should located 
at $E=\mu$ for the original Hamiltonian $H$.
Thus we can describe
non-zero energy states algebraically by suitably choosing $\mu$. 
This is trivial but useful as we demonstrate in Secs.3 and 4.

While we can normalize the molecular orbitals as ${\psi} ^\dagger {\psi}=1$, they overlap with each other in general, 
which implies 
  the anticommutation relation 
$
  \acm{C_m}{C_{m ^\prime } ^\dagger } 
  $ 
  may not be simple. 
We {
  further} decompose the Hamiltonian as
\begin{eqnarray*}
  H-\mu {\cal N} &=& \bm{c} ^\dagger h\bm{c},
\qquad  h = \sum_{m=1}^M {\cal E}_m P_m,
\end{eqnarray*}
where $P_m={\psi}_m{\psi}_m ^\dagger $ is a projection operator with 
$P_m^2 
=P_m$. For non-orthogonal MO's we have
$P_mP_{m ^\prime }\neq 0$ { 
  when the different MO's $m$ and $ m ^\prime $ have a non zero  overlap}.
Since $P_m$'s span a one-dimensional space, the dimension of the Hamiltonian $h$ is at most $M$.  
We still express the Hamiltonian as an $N\times N$ matrix, this should be redundant, namely 
the $(N-M)$-dimensional subspace has to be null, with 
$h$ having $N-M$ zero eigenvalues.
  We can explicitly show this by writing
the $N\times N$ matrix $h$ as  (see footnote 4 of the Ref.\cite{ZQ})
\begin{eqnarray*}
 \PFmat{h} &=&  \Pket{\Psi }\PHmat{{\cal E} }\PTket{\Psi ^\dagger },\
\Pket{\Psi }=\PFvC{\psi_1}\PFvC{\cdots}\PFvC{\psi_M},\
\end{eqnarray*}
where ${\cal E} ={\rm  diag}\,( {\cal E}_1,\cdots,{\cal E}_M)$
is an $M$-dimensional diagonal matrix and $\Psi$ is an $N\times M$ matrix
composed of the molecular orbitals ($\psi$'s) as columns.  
Then, by a simple algebra, the secular equation for $h$ 
becomes
\begin{eqnarray*}
\fl\qquad  \det\nolimits_N (\lambda E_N- h) &=&
\lambda ^N    \det\nolimits_N ( E_N- \lambda ^{-1}\Psi{\cal E}\Psi ^\dagger   )
=  \lambda ^{N-M}    \det\nolimits_M ( \lambda E_M- {\cal E}\Psi ^\dagger\Psi   ) =0,
\end{eqnarray*}
where $E_{n}$ is an $n$-dimensional unit matrix.  
Then one can see that $h$  has  $(N-M)$-fold 
degenerate zero eigenvalues $\lambda =0$.  
These {
  zero-energy states} are topological in 
that they are stable against
continuous deformations of the parameters such as
${\cal E}_m$'s and $\psi_{i m}$'s.
These zero states are stable 
as far as the number of the molecular orbitals is fixed.  It is a finite dimensional analogue of
the Atiyah-Singer's index theorem\cite{ST}.

Following the idea, 
we state the theorem in a slightly extended form, which we shall 
use later in the present paper.   
For the projection $P_m=P_m^2=P_m ^\dagger  $, let us define its dimension,
{
  ${\rm dim}\, P_m= {\rm Tr}P_m ={\rm rank}\, P_m$.}
Since 
the projection operator has eigenvalues $0$ or $1$,
{
  the dimension is  a number
of nonzero eigenvalues of $P_m$,
  which  also  coincides with the rank of the matrix represerntation of $P_m$}. 
Then the number of zero modes, $Z$, should satisfy a condition,
\begin{eqnarray*}
  Z & \ge  &N-\sum _m {\dim }\, P_m.
\label{dimension}
\end{eqnarray*}

We note here that the flat bands at the zero energy
of the chiral symmetric models\cite{ST,DF86,Lieb,ShimaAoki,HAnjp}
have been discussed with the argument presented above, 
where a square of the Hamiltonian is considered.  
Since the hamiltonian in the chiral
class is written as $\mmat{O}{D}{D ^\dagger }{O}$
in a suitable basis, 
its square is $\mmat{D D ^\dagger }{O}{O}{D ^\dagger D}$.
When $D$ is an $N\times M$ matrix ($N>M$),
one may identify $\Psi=D$ and ${\cal E}=E_M $.
The $(1,1)$ block of this squared Hamiltonian, $D D ^\dagger  $, corresponds
to the present hamiltonian. Then the theorem here
guarantees the existence of zero-energy states of the chiral symmetric Hamiltonian 
with degeneracy  $N-M$.
We can also note that the overlapping molecular orbitals in real space are 
discussed in the context of the rigorous treatment
of the ferromagnetism on the
Hubbard model\cite{Lieb,MT,line}. Non particle-hole symmetric
flat bands on special shape of the lattice (partial line graphs)
are discussed as well\cite{miya}.
Then the relation to the present analysis
can be an interesting problem, and should be discussed in  the future.

\section{Weaire-Thorpe model}
\label{sec:wt}
Weaire and Thorpe considered a simple but multi-orbital
tight-binding model for the $sp^3$ electrons
on the diamond lattice\cite{WT}, where the original motivation 
was to treat amorphous silicone.  
Let us reproduce the model here for later references.  
We start with $sp^3$-hybridized orbitals 
on a single tetrahedron.  
The local Hamiltonian for the tetrahedron reads 
\begin{eqnarray*}
  H_{sp3} &=&   \epsilon _s c_{s} ^\dagger c_{s}
  +\epsilon _p (
  c_{p_x} ^\dagger c_{p_x}
  + c_{p_y} ^\dagger c_{p_y}
  + c_{p_z} ^\dagger c_{p_z})
  \equiv \bm{c} ^\dagger h_{sp^3} \bm{c} ,
\end{eqnarray*}
where  $c_i \;\;(i= s, p_x, p_y, p_z)$ is an annihilation operator of the bond orbitals 
with energy levels $\epsilon _s$ for the s orbital and $\epsilon _p$ for the p orbitals,  and
\begin{eqnarray*}
  \fl\quad
  \ \
h_{sp^3} =  \epsilon_{p}E_4+V_1
  \left[
    \begin{array}{cccc}
      1 & 1 & 1 & 1 \\
      1 & 1 & 1 & 1 \\
      1 & 1 & 1 & 1 \\
      1 & 1 & 1 & 1 
    \end{array}\right], 
 \ \bm{c} = \mvecfour{c_0  }{c_1  }{c_2 }{c_3  }=
  \frac {1}{2} \left[
    \begin{array}{cccc}
      1 & 1 & 1 & 1 \\
    1 & 1 & -1 & -1 \\
    1 & -1 & 1 & -1 \\
    1 & -1 & -1 & 1
    \end{array}\right]
  \mvecfour{c_s  }{c_{p_x}} {c_{p_y}}  {c_{p_z}}  .
\end{eqnarray*}
Here $E_4$ is the $4\times 4$ unit matrix, and  $V_1= \frac 1 4( \epsilon _s-\epsilon_{p} )$ is proportional to the 
s-p level offset. 
With this bond basis, the WT model 
for the silicon atoms on the diamond lattice considers 
only the hopping, denoted by $V_2$, between the bond-sharing orbitals.
\begin{figure} [h] 
\begin{center} 
\includegraphics[width=18cm]{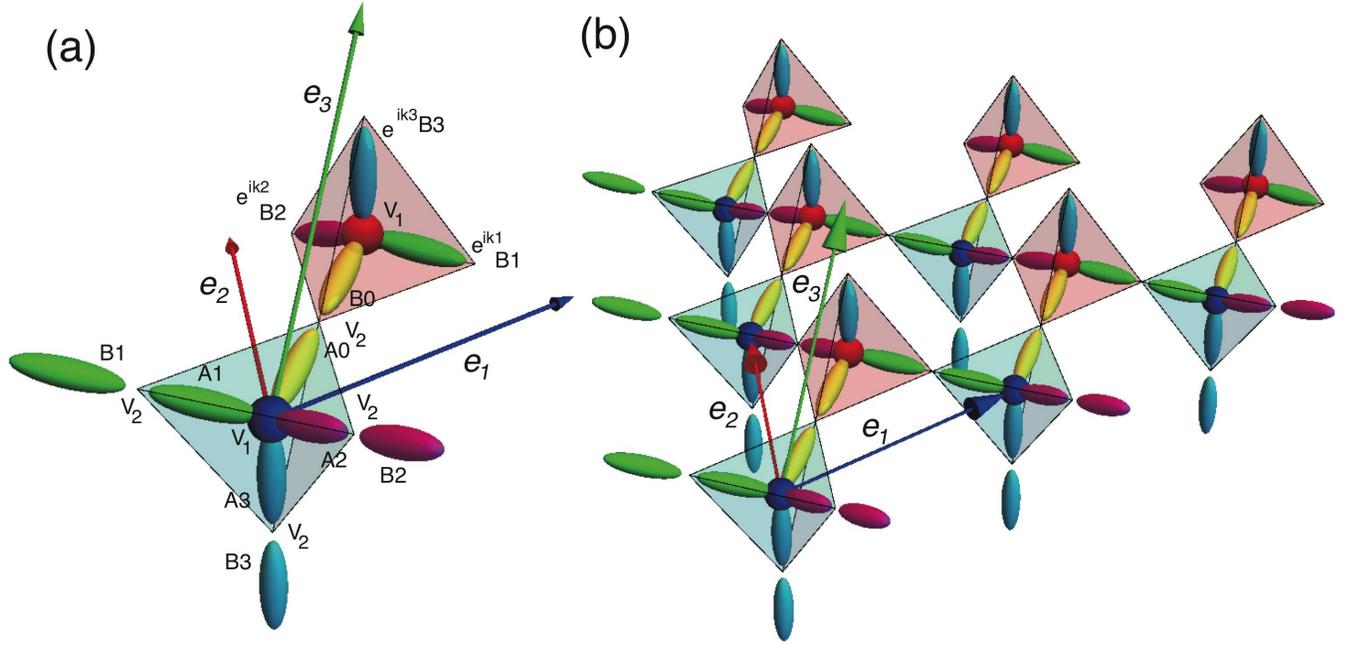}             
\caption{
(a): Unit cell of the WT model for the diamond lattice.
(b)Array of the tetrahedra for the two-dimensional atomic layer (silicene), which has two primitive vectors 
$e_1, e_2$.  If we stack the layer with a primitive vector $e_3$, the WT model in 3D is recovered. 
}
\label{fig:unit}
\end{center} 
\end{figure} 

For the diamond lattice (See Fig.\ref{fig:unit}(a)),
the Hamiltonian in the Bloch picture of the WT model ($H_k$ in the appendix B of \cite{WT}, but note that here we take the origin of energy at $\epsilon_{p} $, so that the energy in \cite{WT} is shifted by $-V_1$ from ours) is written as
\begin{eqnarray*}
\fl\quad  H_{WT}(\bk) =
 \left[ \begin{array}{cccccccc}
     V_1 & V_1 & V_1 & V_1 & V_2 & 0& 0& 0 \\
     V_1 & V_1 & V_1 & V_1 & 0& V_2 &  0& 0 \\
     V_1 & V_1 & V_1 & V_1 & 0&0 &V_2 &   0 \\
          V_1 & V_1 & V_1 & V_1 &  0& 0& 0& V_2 \\
           V_2  & 0 &  0& 0  & V_1  & V_1 e^{i(k_y+k_z)} &V_1 e^{i(k_z+k_x)} & V_1e^{i(k_x+k_y)}  \\
          0& V_2 &  0& 0  &V_1 e^{-i(k_y+k_z)}  & V_1 &V_1 e^{i(k_x-k_y)} & V_1e^{-i(k_z-k_x)}  \\
          0& 0& V_2 & 0 &V_1 e^{-i(k_z+k_x)}  &V_1 e^{-i(k_x-k_y)} & V_1 & V_1e^{i(k_y-k_z)}  \\
          0 & 0& 0& V_2 &V_1 e^{-i(k_x+k_y)}  &V_1 e^{i(k_z-k_x)} &V_1 e^{-i(k_y-k_z)} & V_1
   \end{array}
   \right]
 \\
 \fl\qquad\qquad\quad
 =  
    \mmat{H_V(0)}{V_2 E_4}{V_2 E_4}{H_V(\bk)},
 \end{eqnarray*} 
 \begin{eqnarray*}  
  H_V(\bk)&=& 4V_1 \psi_\bk \psi_\bk ^\dagger,
    \\
    \psi_{\bm{k}} &=&  \frac 1 2 \mvecfour{e^{i(k_x+k_y+k_z)}}{e^{i k_x}}{e^{i k_y}}{e^{i k_z}}
    =   
    \frac 1 2 \mvecfour{1}{e^{-i k_1}}{e^{-i k_2}}{e^{-i k_3}} e^{i(k_x+k_y+k_z)},
\label{WTbasis}
\end{eqnarray*}
where $k_1=k_y+k_z$, $k_2=k_z+k_x$, $k_3=k_x+k_y$, and
$\psi_\bk \psi_\bk ^\dagger \propto H_V$ is a projection onto the space spanned by $\psi_\bk$.
We can then  apply the discussion in sec.{\ref{sec:overlap}} to note that 
$4\times 4$ matrix $H_V(\bk)$  has at most one nonzero energy
that corresponds to a localised molecular orbital at each tetrahedron, while 
there are three zero-energy flat bands (in the present choice of 
the origin of energy).

With this observation, we can introduce 
two representations for
the $8\times 8$ Hamiltonian as 
$H_{WT}(\bk)\pm V_2 E_8$, where
$E_8$ is an $8\times 8$ unit matrix.  
Although one might think the choice of the origin of the energy to be irrelevant, 
the whole point here is: we want to deal with flat bands that have 
nonzero energies.  To do that, we can shift the origin of the energy 
to apply an algebraic argument.  
Now, if we take the plus sign, we have
\begin{eqnarray*}
  \fl\qquad
  H_{WT}(\bk)+ V_2 E_8 &=&
  \mmat{H_V(0)}{0}{0}{H_V(\bk)}
  +
  V_2\mmat{E_4}{E_4}{E_4}{E_4}
  \\
  \fl\qquad  &=& 
  4 V_1 P_1
  +
  4 V_1 P_2
  +
  2 V_2 P_{3+},
\end{eqnarray*}
\begin{eqnarray*} 
  \fl\quad
  P_1 &=& \mmat{\psi_0 \psi_0 ^\dagger}{O}{O}{O}=\Psi_1\Psi_1 ^\dagger,  \    P_2 =\mmat{O}{O}{O}{\psi_\bk \psi_\bk ^\dagger}= \Psi_2\Psi_2 ^\dagger,  \   P_{3+} = \frac 1 2 \mmat{E_4}{E_4}{E_4}{E_4}=
  \Psi_{3+}\Psi_{3+} ^\dagger,
\end{eqnarray*}
\begin{eqnarray*} 
  \Psi_1= \frac {1}{{2} }
 \left[ \begin{array}{c}
    1 \\
    1 \\
    1 \\
    1 \\ 
    0 \\
    0 \\
    0 \\
    0
   \end{array}\right],\quad
 \Psi_2= \frac {1}{{2} }
 \left[ \begin{array}{c}
    0 \\
    0 \\
    0 \\
    0\\
         {1}\\{e^{-i k_1}}\\{e^{-i k_2}}\\{e^{-i k_3}}
   \end{array}\right],\quad
 \Psi_{3+} = \frac 1 {\sqrt{2} } \mvec{E_4}{E_4},
 \end{eqnarray*}
where $P_i$'s are projections with $P_i^2=P_i$ and ${\rm dim} P_i= {\rm Tr}\, \Psi_i\Psi_i ^\dagger =  {\rm Tr}\,  \Psi_i ^\dagger \Psi_i =1,1,4$ respectively 
for $i=1,2,3+$, and 
$\Psi_1 ^\dagger \Psi_1 = \Psi_2 ^\dagger \Psi_2 =1, 
\Psi_{3+} ^\dagger \Psi_{3+} =E_4$. 
If we count the dimensions, $H_{WT}(\bk)+V_2 E_8$ has at most ${\rm dim} P_1+{\rm dim} P_2+{\rm dim} P_{3+}=1+1+4$ nonzero-energy bands,
that is, there are $8-6=2$ zero-energy flat bands.

Similarly we have  
\begin{eqnarray*}
  H_{WT}(\bk)- V_2 E_8 &=&
  \mmat{H_V(0)}{0}{0}{H_V(\bk)}
  -
  V_2\mmat{E_4}{-E_4}{-E_4}{E_4}
  \\
  &=& 
  4 V_1 P_1
  +
  4 V_1 P_2
  -
  2 V_2 P_{3-},
\end{eqnarray*}
\begin{eqnarray*} 
 P_{3-} = \Psi_{3-} (\Psi_{3-} ) ^\dagger,
\end{eqnarray*}
\begin{eqnarray*} 
 \Psi_{3-}= \frac 1 {\sqrt{2} } \mvec{-E_4}{E_4},
\end{eqnarray*}
where $P_{3-}$ is a projection similar to $P_{3+}$.  
Now we see that $H_{WT}-V_2 E_8$ has at most $1+1+4$ nonzero-energy bands,
that is, there are again two zero-energy flat bands.

Hence we end up with two flat bands at each of 
$E=\pm V_2$ in the WT Hamiltonian.  
To compare with Ref.\cite{WT}, we need to shift the energy, after which the flat bands are at $-V_1\pm  V_2$,
since our choice of the $H_{WT}$ is shifted by $-V_1E_8$ from the hamiltonian
in Ref.\cite{WT}.  
An essential point is that we have succeeded in describing 
the {\it flat bands at nonzero energies} 
{\it algebraically} 
in terms of the overlapping molecular orbitals, here 
exemplified in the WT model.

\section{Silicene in a Weaire-Thorpe type model}
Now we come to the original aim at describing silicene.  
We start with an observation that 2D silicene can also be captured in a 
manner similar to the WT model in 3D. 
As indicated  in Fig.\ref{fig:unit}(b), we have three primitive vectors,
$e_1, e_2, e_3$, from which we have three reciprocal vectors for the 3D diamond lattice.  
Corresponding 2D momentum components are given by $(k_1,k_2)$ with $k_3=0$.  
The Hamiltonian can be obtained from that in 3D by cutting the bonds at the blue bonds in Fig.\ref{fig:unit}.
Then the Hamiltonian as a simple extension of the WT model for silicene can be taken as 
\begin{eqnarray*}
\fl\qquad  H_{\rm Silicene}^{\rm 0} = \mmat{H_V(0)}{V_2 E_4^C}{V_2 E_4^C}{H_V(\bk)},\quad
  E_4^C =    \mmat{E_3}{}{}{0}=E_4-{\cal E},\quad
  {\cal E}={\rm diag}\,(0,0,0,1).
\end{eqnarray*}
We have to note that the bonds normal to the
two-dimensional plane, which are originally dangling bonds after 
the dissection but can be treated with hydrogen termination 
for instance, are different from the other bonds.  
A simple way for implementing this is to modify the Hamiltonian into
\begin{eqnarray*}
  H_{\rm Silicene}^{\epsilon_{H}}(\bk) = \mmat{H_V(0)-\epsilon_{H} {\cal E}}{V_2 E_4^C}{V_2 E_4^C}{H_V(\bk)-\epsilon_{H} {\cal E}},
\end{eqnarray*}
where the energy $\epsilon_{H} $ can be controlled by how the 
bond is chemically terminated.

Now, an interesting observation 
is that one can precisely apply the discussion given in Sec.2 to show that the model has flat bands.
This follows from a simple observation that
\begin{eqnarray*}
\fl\qquad  H_{\rm Silicene}^{\epsilon_{H} }+  V_2 E_8 &=& 4V_1 P_1 + 4 V_1 P_2
  + V_2
  \mmat
      {E_4^C}{E_4^C}
      {E_4^C}{E_4^C} +(V_2-\epsilon_{H}) \mmat{{\cal E}}{O}{O}{{\cal E}}
      \\
      \fl\qquad
&=& 4V_1 P_1 + 4 V_1 P_2
  + 2V_2 P_{3+}^C
  +(V_2-\epsilon_{H}) P_5,
\end{eqnarray*}
\begin{eqnarray*}
\fl\qquad  H_{\rm Silicene}^{\epsilon_{H} }-  V_2 E_8 &=& 4V_1 P_1 + 4 V_1 P_2
  - V_2
  \mmat
      {E_4^C}{-E_4^C}
      {-E_4^C}{E_4^C} -(V_2+\epsilon_{H}) \mmat{{\cal E}}{O}{O}{{\cal E}}
      \\
      \fl\qquad
&=& 4V_1 P_1 + 4 V_1 P_2
  - 2V_2 P_{3-}^C
-(V_2+\epsilon_{H}) P_5,
\end{eqnarray*}
where
\begin{eqnarray*} 
  P_{3\pm} ^C =  \frac 1 2 \mmat{ E_4^C}{\pm E_4^C}{\pm E_4^C}{ E_4^C}=
  \Psi_{3\pm}^C(\Psi_{3\pm}^C) ^\dagger,
\\
  \Psi_{3\pm}^C   =\frac 1 {\sqrt{2} }
  \mvec{\pm E_4^C}{E_4^C},
  \\
  (    \Psi_{3\pm}^C ) ^\dagger  \Psi_{3\pm}^C  = E_4^C=(E_4^C)^2,\
   {\rm Tr}  E_4^C=3,
  \\
  P_5 =  \Psi_5\Psi_5 ^\dagger = \mmat{{\cal E}}{O}{O}{{\cal E}}=P_5^2,
  \\
  \Psi_5 = (\Psi_{5,1},\Psi_{5,2}),\   \Psi_{5,1}=\mvec{{\cal E}}{O},\ \Psi_{5,2}=\mvec{O}{{\cal E}},
  \\
  \Psi_{5,1} \Psi_{5,1} ^\dagger  = \mmat{\cal E }{O}{O}{O},\
    \Psi_{5,2} \Psi_{5,2} ^\dagger  = \mmat{O}{O}{O}{\cal E },\
  \\
  \Psi_{5,i} ^\dagger \Psi_{5,i}= {\cal E }^2= {\cal E }  ,\ (i=1,2),\; \Psi_{5,1} ^\dagger \Psi_{5,2}  = \Psi_{5,2} ^\dagger \Psi_{5,1}   =O,
\\
  \Psi_5 ^\dagger   \Psi_5 =
  \mmat
      {\Psi_{5,1} ^\dagger\Psi_{5,1}  }
      {\Psi_{5,1} ^\dagger\Psi_{5,2}  }
      {\Psi_{5,2} ^\dagger\Psi_{5,1}  }
      {\Psi_{5,2} ^\dagger\Psi_{5,2}  }
      =\mmat{{\cal E}}{O}{O}{{\cal E}}.
\end{eqnarray*}
The dimensions of the projections are evaluated as 
\begin{eqnarray*} 
    {\rm dim} P_{3\pm}^C &=&
    {\rm Tr} \Psi_{3\pm}^C {\Psi_{3\pm}^C} ^\dagger
 = {\rm Tr} {\Psi_{3\pm}^C} ^\dagger\Psi_{3\pm}^C   =  {\rm Tr}  E_4^C =      3,
    \\
      {\rm dim}P_5 &=&   {\rm Tr} P_5= {\rm Tr} \Psi_5 \Psi_5 ^\dagger ={\rm Tr} \Psi_5 ^\dagger\Psi_5    =2  {\rm Tr} {\cal E} =      2.
\end{eqnarray*} 

Since we have now expressed the Hamiltonian as a linear combination of projection operators, 
we can see that the wavefunctions associated with
$P_1$ and $P_2$ are localised within each tetrahedron,
the ones with  $P_5$ are localised within the bonds perpendicular to the plane while
the ones with  $P_{3\pm}^C$ extend over the 2D plane.
Then counting the dimensions tells us 
that $H_{\rm Silicene}^{\epsilon_{H} }\pm V_2 E_8$ has at most ${\rm dim}P_1+{\rm dim}P_2+{\rm dim}P_{3\pm}+{\rm dim} P_5 = 1+1+3+2=7
$ nonzero-energy bands,
that is, there is a 8-7=1 flat band at $\pm V_2$ generically.

At a special point of $\epsilon_{H}=\pm V_2 $, 
the coefficient of the projection $P_5$ for the expansion of
$H_{\rm Silicene}^{\epsilon_{H}=\pm V_2 }\pm V_2 E_8 $ 
happen to vanish.  
Then additional two dimensional space of the
Hamiltonian becomes null, which makes 
the flat bands at energies  $\mp V_2$ three-fold degenerate.

As for the signs of the parameters, one has
$V_1<0$ since $\epsilon_{s}<\epsilon_{p}  $, 
and $V_2<0$ since the hopping gains energy.  
Further it is natural to assume the bonds normal to the plane are close to 
be a dangling bond, that is $\epsilon_{H}<0 $ for 
a free-standing silicene.  
Then we may consider the case $\epsilon_{H}=V_2 $ 
belongs to a regime that contains a free-standing silicene. 
Thus we can see algebraically 
that the difference in the structure of the Hamiltonian 
produces the following: while in 3D 
the dispersive bands with Dirac cones are sandwiched 
by doubly-degenerate flat bands, the band structure in 2D silicene
has dispersive bands with cones sandwiched 
by triply-degenerate flat bands and a non-degenerate flat 
band.

\begin{figure} [h] 
\begin{center} 
\includegraphics[width=15cm]{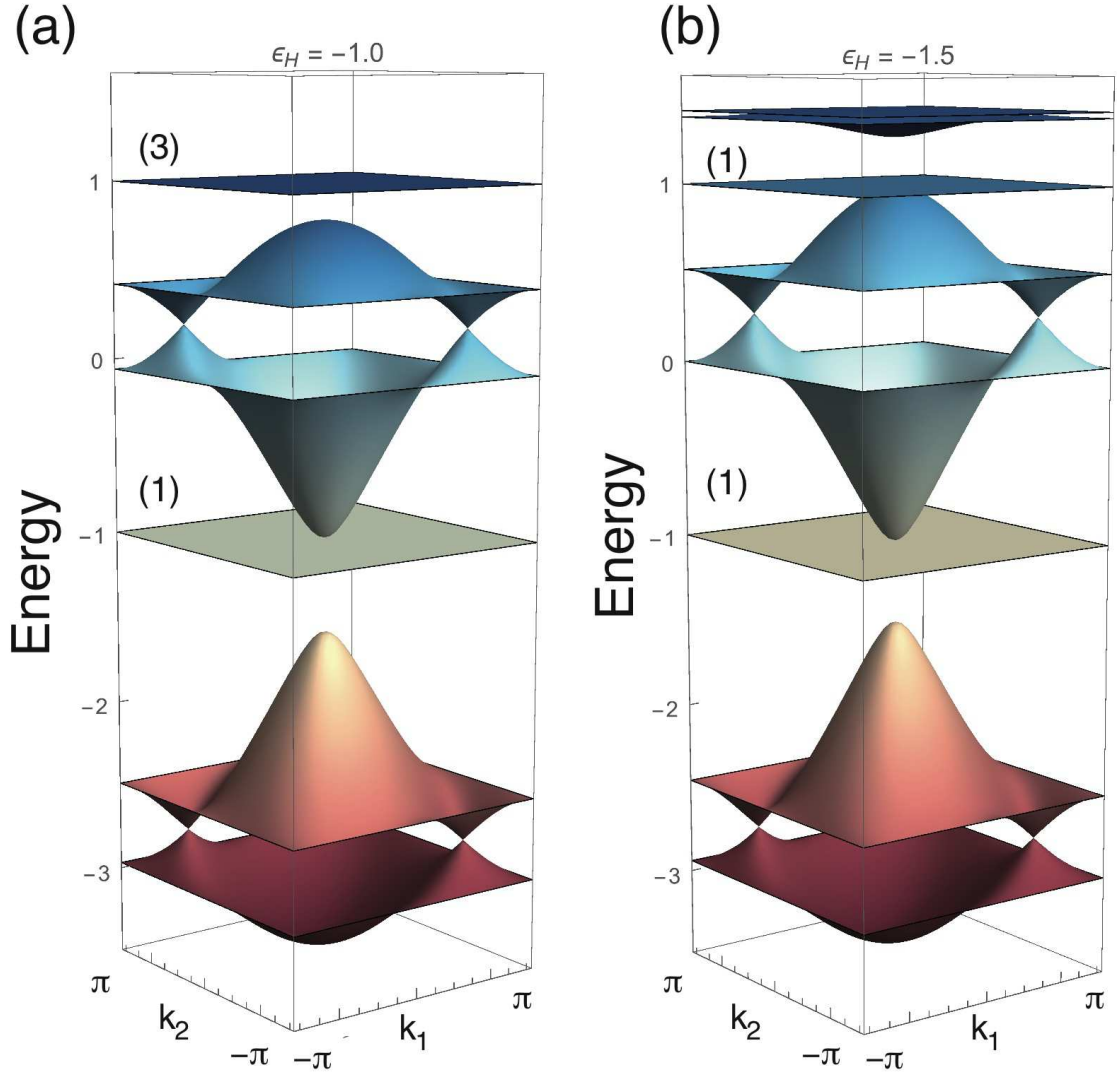}             
\caption{
  Band structure of the present model inspired by 
the WT model for silicene (with $V_1=-0.7$ and $V_2=-1.0$), 
where a region ($-\pi<k_1,k_2<\pi$) 
wider than the first Brillouin zone is displayed, and
the numbers in parentheses indicate the band degeneracy.
  (a) The case of $\epsilon_{H}=V_2=-1 $, for which 
there are  three-fold degenerate
  flat bands at the energy $-V_2(=1)$ and a non-degenerate 
flat band at $V_2(=-1)$.  
{
  (b) A general case with $\epsilon_{H}< V_2 (=-1.0)$.
Here we put $\epsilon_{H}=1.5 V_2 =-1.5$. There are still two
non-degenerate flat bands at $\pm V_2 (=\mp 1)$ touching 
the band that connects to the Dirac cones at $S$ point.}
This is to be compared with Fig.\ref{fig:ts}).
}
\label{fig:disp-sil}
\end{center} 
\end{figure}

\begin{figure} [h] 
\begin{center} 
\includegraphics[width=10cm]{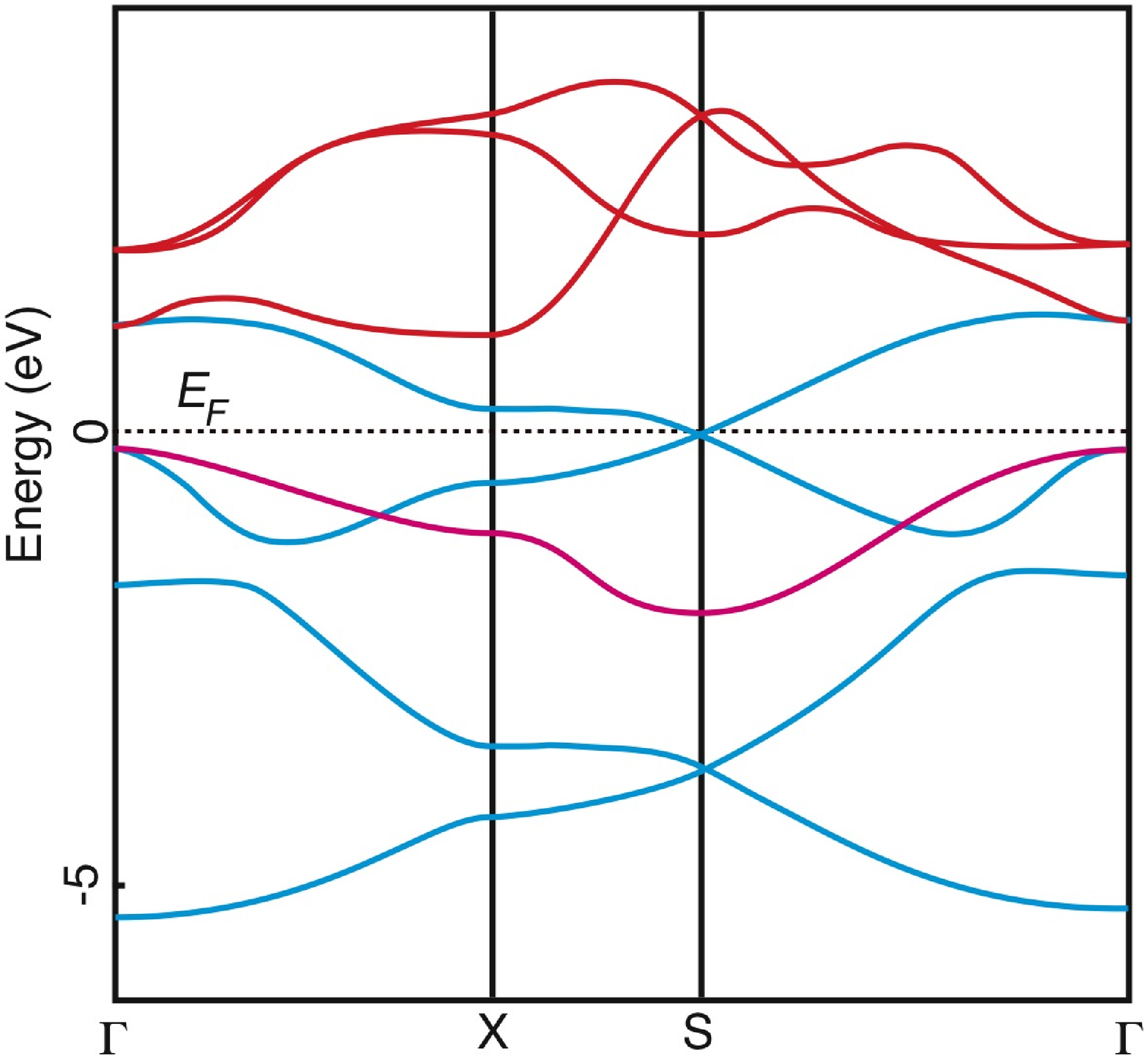}             
\caption{
  Band structure  of silicene obtained by Takeda-Shiraishi.\cite{TS}
  The energy bands that can be adiabatically traced back to the flat bands are shown in red and magenta, while 
the pairs of bands each of which contains a massless Dirac cone are shown in blue.
}
\label{fig:ts}
\end{center} 
\end{figure}

Now let us demonstrate the above analytic formulation 
by numerical results for the band structure 
in Fig.\ref{fig:disp-sil}.  Panel (a) depicts 
the {
  ideal} 
case of {
  $\epsilon_{H}=V_2=-1.0 $, where we can 
indeed see the triply-degenerate flat bands at $E=-V_2=1$ 
along with a non-degenerate flat band at $E=V_2=-1$.  
The result confirms the analytic discussion above. 
Panel (b) depicts a general case of $\epsilon_H<V_2=-1$, where 
we can see that the two out of the three flat bands become
somewhat dispersive  around $E=-\epsilon_{H}$ while 
one flat band remains to be flat at $E=-V_2=1$.
The two nearly-flat bands 
derive from the dangling bonds perpendicular to the plane. }
Interestingly, if we compare the result for the 
dispersion of silicene obtained by Takeda and 
Shiraishi\cite{TS} as reproduced in Fig.\ref{fig:ts} here, 
we can see that they roughly agree with each other 
in terms of the bands'
{
  widths, ordering  and multiplicity}.  
Specifically, we can make the following observation. 
In both of the realistic band structure\cite{TS} and 
the present algebraic treatment for silicene, 
{
  we have basically two bands that contain Dirac dispersions.
  The  bands are 
sandwiched by two narrow (ideally flat) 
bands, each of which touches the dispersive ones 
at $\Gamma$ point.
Above these, 
there are two nearly-flat bands\cite{TS}.
Thus the three narrow bands above the Dirac cone can be traced 
back to the three-fold degenerate
  flat bands in the idealised model. 
  Of course, this is only a rough mapping, but these qualitative
  agreements in the band structure between the WT-like model 
and the realistic band calculation gives an insight into 
a physical origin  of the electronic band structure of silicene.
}
The existence of the (nearly) flat bands 
has an intuitive origin as well. 
If electrons
hop as molecular orbitals, 
this imposes a strong constraint 
on the description in terms of the original electron. 
Namely, the degrees of freedom other than the molecular-orbital hopping remain 
more or less frozen, which form the nearly-flat (or less dispersive) bands.

Within the present simple model, 
one may expect the electrons at the perpendicular bonds can be  stabilized
by the hydrogen termination.
It can be modeled as $\epsilon_{H}=-V_2 >0$.
{
  Then the flat bands originally situated at $-V_2>0$ will 
move down to $V_2$, that is, below the energy region of
the Dirac cones.
Then the Fermi energy, 
which is originally situated 
at the Dirac point since the $sp^3$-bands are half-filled 
in silicene, will move away from the original Dirac point.
}

\subsection*{Dirac cones and effects of buckling}
We can even extend the present argument when there is some buckling 
in the structure.  For this we start with an observation that 
existence of the Dirac cones in silicene 
is due to the time-reversal ($T$) and {
  crystal symmetry (e.g. reflection, $R$, 
that exchanges bonds 1 and 2)\cite{Wa,Lo,He}.}
This is readily seen in the present model, 
where the momentum dependence is expressed in terms of 
$\psi_\bk$. At K and K$^\prime $ points, 
$\psi_\bk$ is expressed and transformed as
\begin{eqnarray*}
  \psi_\bk &=& \frac 1 2 \mvecfour{1}{\omega }{ \omega ^2} {1},\quad T
  R\psi_\bk =
  \frac 1 2 T \mvecfour{1}{ \omega ^2}{\omega }{1}=
  \psi_\bk,
\end{eqnarray*}
with $\omega\ne 1 $ being a nontrivial cubic root of unity, 
{
  which is the origin of the degeneracy 
at K and K$^\prime $. }
Since the degeneracy is lifted at the general momentum, 
the energy gap is linearly vanishing (Dirac cones) around K and K$^\prime $. 
The property is stable as far as the symmetries $T$ and {
  $R$}
remain.

Then what are the effects of the buckling in the $sp^3$ structure observed in silicene? 
Since the main effect of the buckling 
is a modification of the bond-to-bond angles 
among the $sp^3$ orbitals, 
the local hamiltonian within a tetrahedron depends on the buckling angle $\theta $ as
\begin{eqnarray*}
  h_{\rm local}(\theta) &=& \sum_{\langle  i,j \rangle }V_{ij}c_i ^\dagger c_j
+{\rm h.c.},
  \\
  V_{ij} &=&
  \mchss
      {V_1}{(\langle i,j \rangle =\langle 01 \rangle,\langle 12 \rangle,
        \langle 20 \rangle)
      }
      {V_1 ^\prime }{(\langle i,j \rangle =\langle 0 3 \rangle,\langle 1 3\rangle,\langle 2 3 \rangle)
      }
      \\
     \frac { V_1 ^\prime}{V_1}  &=& \frac {\cos\theta }{\cos\theta_0 } , \ \cos\theta _0 = -\frac {1}{3} ,
\end{eqnarray*}
where $c_3$ is the annihilation operator for the bond that is perpendicular
to the silicene plane.
{
  Since the two bonds coupling 
  the neighboring tetrahedra 
  remain straight 
even with the  buckling, the inter-site term $V_2$ is not modified.}
As far as the buckling is small, the effects can be
considered by a Hamiltonian,
\begin{eqnarray*}
  H_{\rm Silicene}^{\epsilon_{H},\cos\theta }(\bk) &=&  \mmat{H^{\theta }_V(0)-\epsilon^\theta _{H} {\cal E}}{V_2 E_4^C}{V_2 E_4^C}{H^\theta _V(\bk)-\epsilon^\theta _{H} {\cal E}},
  \\
  H^\theta _V(\bk)&=&4 V_1 \psi^\theta_\bk (\psi^\theta_\bk ) ^\dagger,
  \\
  \psi^\theta_\bk &=& {\rm  diag}\,
  \left(\frac {\cos\theta  }{\cos\theta_0 },1,1,1 \right) \psi_\bk,
  \\
  \epsilon^\theta _{H} &=&  \bigg(\frac {\cos\theta}{\cos\theta_0}\bigg)^2
  \epsilon _{H}. 
\end{eqnarray*}
This implies the Dirac cones and the flat bands we 
have focused are stable against the buckling. 
This is also regarded as an aspect of the topological 
stability.  
When the buckling becomes large enough, however, 
$sp^2$-character of the hybridized orbitals will generate 
direct hoppings among the out-of-plane bonds, 
which is not included in the present formalism.

\section*{Conclusion}
In the present paper, after introducing a generic argument for flat bands, 
the Weair-Thorpe model, originally conceived for 3D silicon, is extended to 2D silicene.  A surprise revealed here is that 
the flat bands that arise in the WT model for the hybridized $sp^3$ orbitals in 3D 
also appear, in ideal situations, in silicene, but with different degeneracies in the flat bands for an 
algebraic reason.  
In this picture, the band structure, including the flat ones, 
are crucially controlled by the out-of-atomic-plane 
orbits.  We have further pointed out there are pairs 
of bands each of which contains a Dirac cone.

The flat bands emerging in our treatment 
can be theoretically interesting and important as a 
multi-orbital effect\cite{ShimaAoki}.  
Finite samples will accommodate characteristic edge 
states.  
The flat bands in the idealised model, which should become 
dispersive in realistic situations, will still have large 
density of states.  If we can shift the chemical potential close to 
the flat bands, e.g., by chemical doping, interesting phenomena 
are expected.  Among these are (i) structural instabilities, 
such as those observed experimentally, and (ii) flat-band ferromagnetism\cite{KusakabeAoki}. 
We believe that these will 
help for the synthesis and characterisation of silicene.

\section*{Acknowledgments}
This work was supported in part by JSPS Grant Numbers 26247064(YH,HA), 25610101(YH), 25107005(HA)
and 25610101 from MEXT(YH,HA).

\end{document}